\documentclass[preprint,aps,pre,longbibliography]{revtex4-1}% reprint 
\usepackage{natbib}
\usepackage{graphicx}
\usepackage{subfigure}
\usepackage{amssymb}
\usepackage{amsfonts}
\usepackage{amsmath}
\usepackage{amsbsy}
\usepackage{mathrsfs} 
\usepackage{color}
\usepackage{siunitx}
\usepackage{gensymb} 
\usepackage{soul}%\st{...}
\usepackage{lineno}
\usepackage{esint}
\usepackage{mymacros}
\usepackage{soul}

\begin{document}

\title{Spontaneous dynamics of two-dimensional Leidenfrost wheels}
\author{Rodolfo Brand\~{a}o}
\author{Ory Schnitzer}
\affiliation{Department of Mathematics, Imperial College London, SW7 2AZ London, UK}

\begin{abstract}
Recent experiments have shown  that liquid Leidenfrost drops levitated by their vapor above a flat hot surface can exhibit symmetry-breaking spontaneous dynamics (A. Bouillant \textit{et al.}, \textit{Nature Physics}, \textbf{14} 1188--1192, 2018). Motivated by these observations, we theoretically investigate the 
translational and rotational dynamics of Leidenfrost drops on the basis of a simplified two-dimensional model, focusing on near-circular drops small relative to the capillary length. The model couples the equations of motion of the drop, which flows as a rigid wheel, and thin-film equations governing the vapor flow, the profile of the deformable vapor-liquid interface and thus the hydrodynamic forces and torques on the drop. In contrast to previous analytical models of Leidenfrost drops levitating above a flat surface, 
which predict only symmetric solutions, we find that the symmetric Leidenfrost state is unstable above a critical drop radius: $R_1$ for a free drop and $R_2>R_1$ for an immobilized drop. In these respective cases, symmetry breaking is manifested in supercritical-pitchfork bifurcations into steady states of pure rolling and constant angular velocity. In further qualitative agreement with the experiments, when a  symmetry-broken immobilized drop is suddenly released it initially moves at an acceleration $\alpha g$, where $\alpha$ is an angle characterizing the slope of the liquid-vapor profile and $g$ is the gravitational acceleration; moreover, $\alpha$ exhibits a maximum with respect to the drop radius, at a radius increasing with the temperature difference between the surface and the drop. 
\end{abstract}

%\date{\today}

\maketitle

\section{Introduction}
A liquid drop can levitate above a hot surface if the temperature of the surface sufficiently exceeds the boiling temperature of the liquid \cite{Biance:03}. This effect was first studied by J.~G.~Leidenfrost in 1756 and ever since has been a source of great scientific curiosity. The Leidenfrost effect is associated with a sharp transition from nucleate to film boiling as the surface temperature is increased past the so-called Leidenfrost point. The vapor film formed by evaporation at the bottom of the drop prevents direct contact. As a consequence, Leidenfrost drops exhibit increased lifetimes and spectacular mobility,  attributes that either hold promise or are concerning for many applications. Experimental and theoretical studies over the last decade have unraveled the wealth of interesting dynamics exhibited by Leidenfrost drops, including oscillations, bouncing, take-off and directed propulsion using asymmetrically structured surfaces and thermal gradients \cite{Quere:13}.

Surprisingly, recent experiments have revealed that Leidenfrost drops levitating above a flat substrate can also exhibit spontaneous translational and rotational motion  \emph{in the absence of external gradients or geometrical asymmetries} \cite{Bouillant:18}.
In these experiments, a rotational flow is spontaneously established within an immobilized mm-scale Leidenfrost drop --- not too small or too large; once released, the drop begins to move horizontally at an acceleration approximately equal to the product of the gravitational acceleration $g$ and the angle measured between the surface and the liquid-vapor interface, which is asymmetrically deformed.

The nature of the observed spontaneous motion is explored in \cite{Bouillant:18} through measurements of the drop acceleration, interface inclination, temperature distribution and visualization of the internal liquid flow and drop shape. With increasing drop radius, the measured inclination angle symmetrically bifurcates from zero at a first critical radius, growing in magnitude up to a maximum value only to return to zero at a second critical radius; the first critical radius is small relative to the capillary length (near-spherical drops). As already mentioned, the acceleration of an  immobilized drop that is suddenly released is approximately proportional to the inclination angle; it therefore follows similar trends. These findings strongly suggest that the observed spontaneous motion is associated with a symmetry-breaking instability, thus shedding new light on the spectacular mobility of Leidenfrost drops \cite{Bouillant:18:gallery,*Miller:18}. 

We note that asymmetric deformation of the liquid-vapor interface was incidentally observed in earlier studies of immobilized Leidenfrost drops \cite{Celestini2:12,Burton:12}. These studies, however, were focused on the morphology 
of the liquid-vapor interface and did not reveal the bifurcation picture described above, nor any connection between the observed asymmetry and 
 spontaneous dynamics.

Theoretical analyses of 
Leidenfrost drops
\cite{Pomeau:12, Sobac:14,Sobac:17}, as well as related configurations of levitated drops \cite{Duchemin:05,Snoeijer:09}, have traditionally employed a lubrication approximation to model the vapor film, which is either patched or matched to a hydrostatic model of the top side of the drop. These analyses, however, which all neglect the liquid flow based on 
the high liquid-vapor viscosity contrast, predict neither symmetry breaking nor spontaneous dynamics.

In this Rapid Communication, we theoretically describe for the first time the mechanics of symmetry breaking and spontaneous dynamics of Leidenfrost drops based on a simplified two-dimensional model. We focus on the regime of small near-circular drops, in which case it turns out to be straightforward to account for the internal liquid flow. In \S\ref{sec:model}, we formulate the problem and develop the model. In \S\ref{sec:results} we discuss the predictions of the model against the  experiments in \cite{Bouillant:18}. We make concluding remarks in \S\ref{sec:conc}.

\section{Two-dimensional model}\label{sec:model}
Consider a two-dimensional liquid drop (area $\pi R^2$, density $\bar\rho$, viscosity $\bar{\mu}$) levitated by a thin vapor layer (thermal conductivity $k$, density $\rho$, viscosity $\mu$) above a flat solid surface of temperature $\Delta T$ relative to the drop. Following standard modeling of Leidenfrost drops, we assume an isothermal drop at boiling temperature and that area variations due to the evaporation are negligible on the time scale of interest.

A key assumption is $R\ll l_c=\sqrt{\gamma/\bar{\rho}g}$, the capillary length, wherein $\gamma$ is the liquid-vapor interfacial tension; equivalently, the Bond number 
$B=(R/l_c)^2\ll1$. Similarly to the case of a sessile nonwetting drop, in that limit the shape of a two-dimensional Leidenfrost drop is approximately a circle except close to a small ``flat spot'' at the bottom of the drop. The liquid pressure is approximately uniform and equal to the capillary pressure $\gamma/R$.  If that is also the scaling of the  pressure in the vapor film below the flat spot, a balance with gravity gives the scaling $L=BR$ for the flat-spot length. This standard argument assumes that the dynamical stresses associated with the internal liquid flow are $\ll\gamma/R$.

\begin{figure*}[t!]
\begin{center}
\includegraphics[scale=0.49]{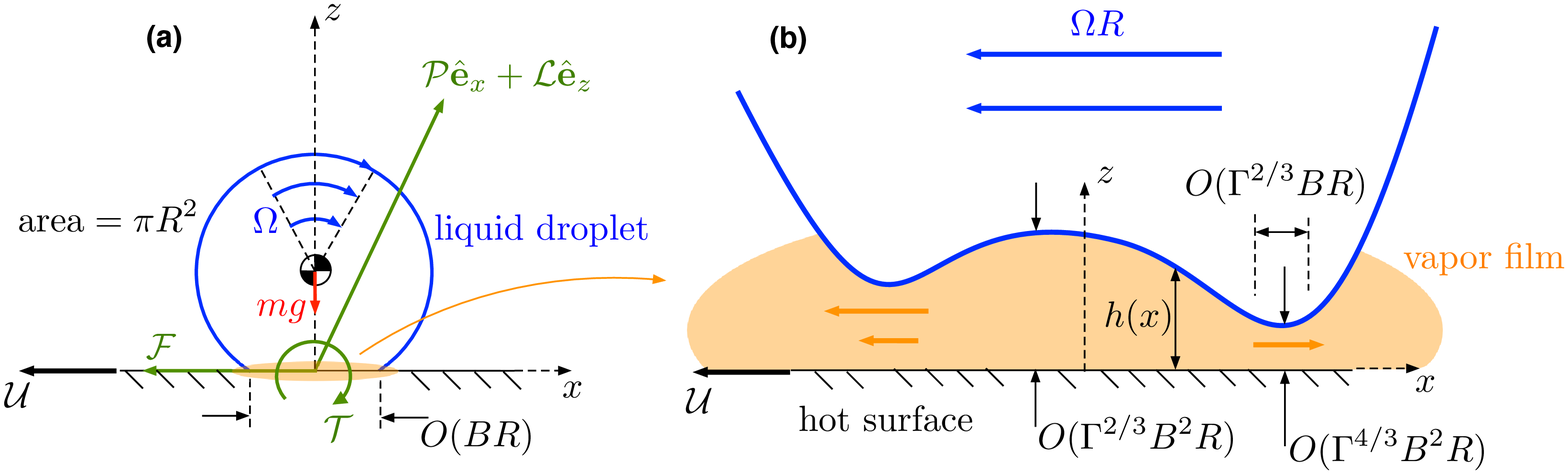}
\caption{Two-dimensional model of a near-circular ($B\ll1$) Leidenfrost drop in a frame co-moving with its center of mass. (a) Drop-scale dynamics. (b) Lubrication region. The small-$\Gamma$ scalings indicated in (b) hold, up to logarithmic factors, in the stationary-symmetric state and close to the onset of spontaneous dynamics.} 
\label{fig:schematic}
\end{center}
\end{figure*}

\subsection{Liquid flow}

Consider now the internal liquid flow, which is usually  thought to be shear-driven by the vapor being squeezed symmetrically outwards from beneath the flat spot. For $B\ll1$, and given the high liquid-vapor viscosity contrast, such a shear-driven flow would be confined to the $O(L)$ vicinity of the flat spot, symmetric and weak relative to the vapor flow \cite{Quere:13}.  We shall see, however, that when the stationary-symmetric state of a Leidenfrost drop is unstable, the net forces and torques exerted on the drop by the vapor film can inertially drive an asymmetric drop-scale flow. The latter flow, in turn, entrains the vapor flow, which is no longer driven solely by evaporation. 

A rotational drop-scale flow is indeed observed in the experiments \cite{Bouillant:18}. Theoretically, given the circular shape of the drop boundary, we hypothesize that the drop-scale flow is a superposition of a rigid-body translation, at velocity $\mathcal{U}(t)\be_x$, and rigid-body rotation, at angular velocity $\Omega(t) \be_y$ (see Fig.~\ref{fig:schematic}a). Here $t$ is time and we introduce unit vectors $\be_x$ and $\be_z$ pointing horizontally and vertically upwards, respectively, with $\be_y = \be_z \boldsymbol{\times} \be_x$.
This ansatz is inspired by the analysis of a  nonwetting drop rolling down a gently inclined plane \cite{Mahadevan:99, Yariv:19:2Drolling,Schnitzer:20}, where a similar leading-order drop-scale solution holds. In the present problem, where we allow for a time dependence of the rigid-body motion, substitution into the time-dependent Navier--Stokes equations and the interfacial conditions on the circular boundary furnishes the condition that 
$\Omega T\gg1$, where $T$ is the inertial time scale on which $\mathcal{U}(t)$ and $\Omega(t)$ vary;
this is in addition to said condition that drop-scale dynamical stresses are $\ll\gamma/R$ \footnote{Relative to the capillary stress $\gamma/R$, the inertial and viscous dynamic stresses in the liquid respectively scale like the Weber number $\mathrm{We}=\bar{\rho} U^2 R/\gamma$ and $B\mathrm{Ca}$, wherein $\mathrm{Ca}=\bar{\mu} U/\gamma$ is the capillary number and $U$ the characteristic velocity defined in \S\S\ref{ssec:scalings}. (The $B$ factor is because the viscous stress vanishes for a rigid-body motion.) We require these dynamic stresses to be $o(B\gamma/R)$, rather than $o(\gamma/R)$, for the reason mentioned below \eqref{vforce}. Furthermore, the condition $1/\Omega\ll T$, or $R/U \ll T$, is equivalent to $\Gamma B^2\ll \mathrm{We}$, wherein $\Gamma$ is defined in \eqref{Gamma def} and we use the estimates for $U$ and $T$ derived in \S\S\ref{ssec:scalings}. We accordingly assume that  $\Gamma B^2\ll \mathrm{We}\ll B$ and $\mathrm{Ca}\ll1$.}.

\subsection{Vapor film}
With this drop-scale physical picture, we turn our attention to modeling the vapor film. We shall employ the lubrication approximation on the premise, to be verified, that the film is thin relative to its $O(L)$ width.
 It is convenient to work in a frame co-moving with the drop center-of-mass, with the corresponding Cartesian coordinate system defined in Fig.~\ref{fig:schematic}.

We denote by $\bu=u\be_x+w\be_z$ the vapor velocity field in the co-moving frame, by $p$ the corresponding vapor pressure field and by $h$ the film-thickness profile. In the lubrication approximation, $p$ is independent of $z$ and the momentum and continuity  equations respectively become
\refstepcounter{equation}
$$
\label{lub eq}
-\pd{p}{x}+\mu\pd{^2u}{z^2}=0, \quad \pd{u}{x}+\pd{w}{z}=0.
\eqno{(\theequation{\mathrm{a},\mathrm{b}})}
$$
At the solid substrate, the flow satisfies no-slip,
\refstepcounter{equation}
$$
\label{bc solid}
u=-\mathcal{U}, \quad w=0 \quad \text{at}  \quad z=0.
\eqno{(\theequation{\mathrm{a},\mathrm{b}})}
$$
To formulate the boundary conditions on the liquid-vapor interface, we assume that (i) the liquid velocity and pressure at the bottom of the drop are approximately uniform and equal  to $-\Omega(t) R \be_x$ and $\gamma/R$, respectively 
\footnote{The deviation of the drop shape from a circular shape, due to gravity, implies a negligible $O(B\Omega R)$ velocity perturbation \cite{Yariv:19:2Drolling}. Meanwhile, a tangential-stress balance at the liquid-vapor interface implies a velocity perturbation $O(\mu L/\bar{\mu}H)$ relative to the characteristic vapor velocity $U$, where $H=\Gamma B^2R$ and $\Gamma$ is defined in \eqref{Gamma def}. For $U=O(\Omega R)$ this implies the condition $\mu/\bar{\mu}\ll B\Gamma$. Flow measurements at the drop base confirm an approximately uniform flow \cite{Bouillant:18}.};
(ii) evaporation at the interface contributes a normal vapor speed $-\lambda/h$, where $\lambda=k\Delta T / \rho l$, $l$ being the latent heat of evaporation \cite{Pomeau:12}; and (iii) the interface velocity $dh/dt$ is negligible compared to the normal vapor speed
\footnote{In the co-moving frame, $h$ varies on the inertial time scale $T$. Using the estimates derived in \S\S\ref{ssec:scalings}, the ratio between $d{h}/dt$ and the normal vapor speed is $B R/U T$, which is small since $B\ll 1$ and $T \gg R/U$.}. 
The kinematic boundary conditions and continuity of tangential velocity at the interface then give
\refstepcounter{equation}
$$
\label{bc h}
u=-\Omega R, \quad w=-\frac{\lambda}{h}-\Omega R \pd{h}{x} \quad \text{at} \quad z=h,
\eqno{(\theequation{\mathrm{a},\mathrm{b}})}
$$
while the dynamic boundary condition at the interface gives
\begin{equation}\label{dyn}
\frac{\gamma}{R}-p=\gamma\pd{^2h}{x^2}.
\end{equation}

Integrating the momentum equation (\ref{lub eq}{a}) together with the boundary conditions (\ref{bc solid}{a}) and (\ref{bc h}{a}), we find 
\begin{equation}\label{u sol}
u+\mathcal{U}=\frac{1}{2\mu}\pd{p}{x}z(z-h)+\frac{z}{h}(\mathcal{U}-\Omega R).
\end{equation}
Next, integrating the continuity equation (\ref{lub eq}{b}) using conditions (\ref{bc solid}{b}) and (\ref{bc h}{b}), we find
\begin{equation}\label{reynolds}
\pd{}{x}\left(\frac{h^3}{12\mu}\pd{p}{x}\right)+\frac{\mathcal{U}+\Omega R}{2}\pd{h}{x}=-\frac{\lambda}{h}.
\end{equation}

To close the thin-film problem governing $p$ and $h$ we require four boundary conditions. These can be represented in terms of coefficients in large-$|x|$ expansions of $h$. Thus, since $p\to0$ in that limit, \eqref{dyn} implies 
\begin{equation}\label{dim h far}
h \sim  \frac{x^2}{2R}+ a_{\pm}x + O(1) 
\quad \text{as} \quad x\to\pm\infty.
\end{equation}
The leading quadratic terms provide two boundary conditions; the correction linear terms provide two more, assuming that the  coefficients $a_{\pm}$ can be determined. One relation between $a_{\pm}$ can be deduced from the vertical force balance 
\begin{equation}\label{vforce}
\mathcal{L}-mg = 0, \quad \mathcal{L}=\int_{-\infty}^{\infty} p \,dx,
\end{equation}
where $m=\pi \bar{\rho}R^2$ is the drop mass per unit length and 
$\mathcal{L}$ the ``lift'' force per unit length: substituting \eqref{dyn} into \eqref{vforce} gives $a_- - a_+=\pi B$. To obtain a second relation, we note that the linear terms in \eqref{dim h far} match with $O(BR)$ drop-scale deviations from a circular shape; such deviations are hydrostatic [15] and therefore even in $x$. It follows that $a_+=-a_-=-\pi B /2$.

\begin{figure}[t!]
\begin{center}
\includegraphics[scale=0.5]{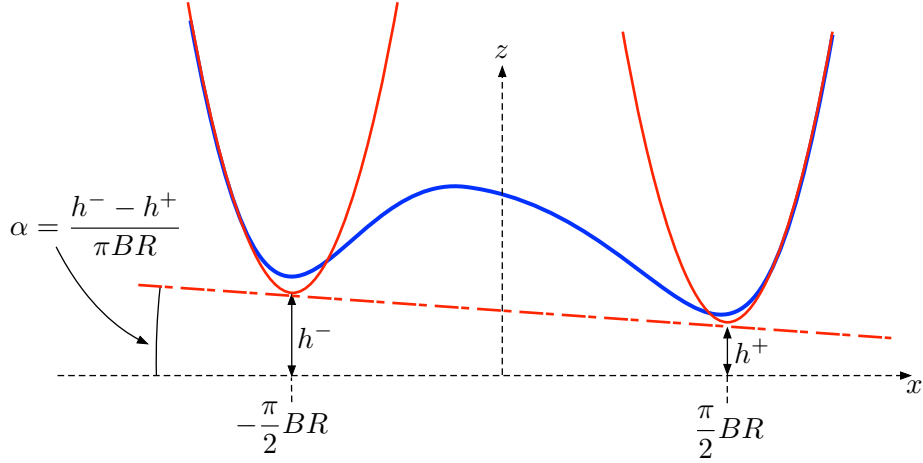}
\caption{The slope $\alpha$ defined in \eqref{alpha def} based on the large-$|x|$ expansions of the thickness profile $h(x)$.}
\label{fig:alpha}
\end{center}
\end{figure}

The lubrication model is now closed and provides $p$ and $h$ for an instantaneous value of the sum $\mathcal{U}+\Omega R$. Then $u$ follows from \eqref{u sol}. It is insightful to rewrite and extend the far-field expansions \eqref{dim h far} in the form
\begin{equation}\label{dim h far expanded}
h\sim \frac{1}{2R}\left(x\mp \frac{\pi BR}{2}\right)^2+h_{\pm} +o(1) \quad \text{as} \quad x\to\pm\infty, 
\end{equation}
where the constants $h_{\pm}$ are outputs of the lubrication problem. The geometric interpretation of these constants can be understood from Fig.~\ref{fig:alpha}. In particular, the slope
\begin{equation}\label{alpha def}
\alpha = \frac{h^--h^+}{\pi B R}
\end{equation}
is small and represents the inclination of the drop base relative to the horizontal surface. Note, however, that $\alpha=0$ does not necessarily imply a symmetric liquid-vapor interface.

\subsection{Drop dynamics}
We turn to the dynamics of the drop as a rigid body. To this end, we consider an integral linear-momentum balance in the $x$ direction and an integral angular-momentum balance in the $y$ direction, with the hydrodynamic forces and torques on the drop calculated based on the lubrication approximation of the vapor film. We thereby find the equations of motion \refstepcounter{equation}
$$
\label{EOM}
m\frac{d\mathcal{U}}{dt}=\mathcal{P}-\mathcal{F}, \quad I\frac{d{\Omega}}{dt}=\mathcal{T}-R\mathcal{P}+R\mathcal{F},
\eqno{(\theequation{\mathrm{a},\mathrm{b}})}
$$
where we define the moment of inertia $I=mR^2/2$, the ``propulsion'' and ``friction'' forces per unit length\refstepcounter{equation}
$$
\label{propulsion-friction}
\mathcal{P}=-\frac{1}{2}\int_{-\infty}^{\infty}p\pd{h}{x}\,dx, \quad \mathcal{F}=\mu(\mathcal{U}-\Omega R)\int_{-\infty}^{\infty}\frac{dx}{h},\eqno{(\theequation{\mathrm{a},\mathrm{b}})}
$$
respectively, and the torque per unit length
\begin{equation}\label{torque}
\mathcal{T}=-\int_{-\infty}^{\infty}xp\,dx.
\end{equation}
Note that the propulsion $\mathcal{P}$ and torque $\mathcal{T}$ are nonlinear functions of $\mathcal{U}+\Omega R$; using \eqref{dyn}, \eqref{dim h far expanded} and \eqref{alpha def}, these can be written 
\refstepcounter{equation}
$$
\label{PT in h}
\mathcal{P}=\frac{1}{2}\alpha m g , \quad \mathcal{T}=\alpha m g R. 
\eqno{(\theequation{\mathrm{a},\mathrm{b}})}
$$

Note that $h$ is invariant under the transformation ${x}\to-{x}$ and $\mathcal{U}+\Omega R\to-(\mathcal{U}+\Omega R)$. It follows that the difference $h_--h_+$, hence  $\alpha$, $\mathcal{P}$ and $\mathcal{T}$ are odd functions of $\mathcal{U}+\Omega R$. Furthermore, the friction $\mathcal{F}$ is a product of the factor $\mathcal{U}-\Omega R$ and a positive even function of $\mathcal{U}+\Omega R$.

\subsection{Scalings}\label{ssec:scalings}
Let $H$ be the characteristic thickness of the vapor film. Assuming the vertical velocity scaling $W = \lambda/H$ based on the evaporation term in (\ref{bc h}b), (\ref{lub eq}{b}) implies the scaling $U=LW/H$ for the horizontal velocity. Comparing the corresponding lubrication pressure scaling 
with the capillary pressure $\gamma/R$ provides the estimate $H = \Gamma B^2R$, where we define the parameter
\begin{equation}\label{Gamma def}
\Gamma=B^{-3/2}(\mu\lambda /\gamma R)^{1/4}\, \propto \,\Delta T^{1/4}R^{-13/4}.
\end{equation}
It follows from \eqref{propulsion-friction} that the horizontal force on the drop is of order $F = \gamma H/R$ and hence that the inertial time scale is of order $T = m U/F$. 

In the present small-$B$ model, $\Gamma$ is the single dimensionless parameter characterizing the drop dynamics \footnote{See Supplemental Material for a dimensionless formulation of the model.}.
We shall see that our model predicts spontaneous dynamics for small values of $\Gamma$. In that case, the vapor film decomposes into a ``bubble'' bounded by two narrow necks, with the scalings corresponding to the symmetric case indicated in Fig.~\ref{fig:schematic}b; these scalings can be derived following \cite{Sobac:14}. The bubble pressure is $\approx \gamma/R$ and hence from the vertical-force balance \eqref{vforce} the distance between the necks is $\approx \pi L$.

\begin{figure}[t!]
\begin{center}
\includegraphics[scale=0.45]{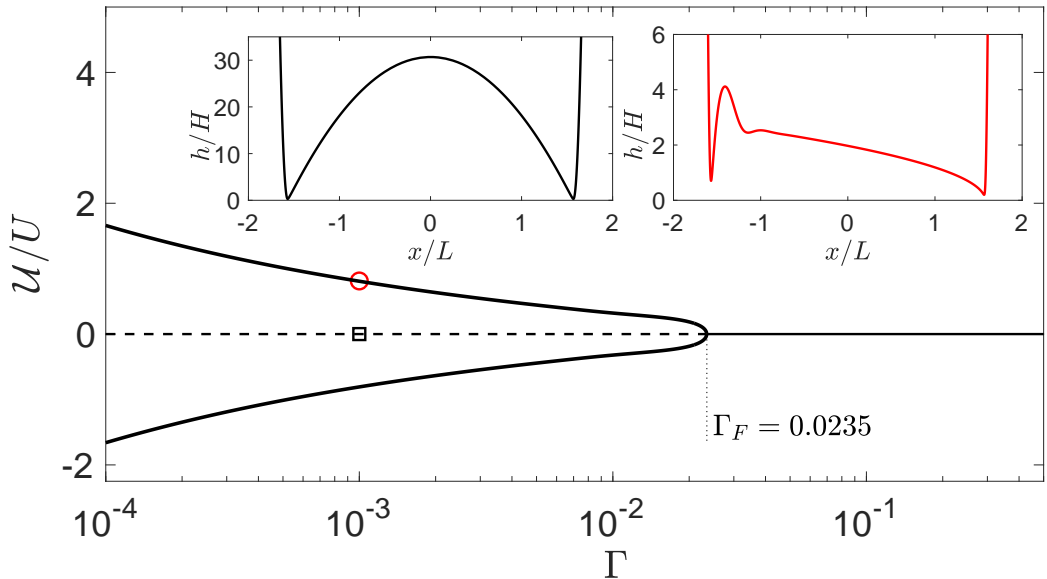}
\caption{Bifurcation diagram for a free drop, whose steady states satisfy the pure-rolling condition $\mathcal{U}=\Omega R$; solid lines --- stable, dashed line --- unstable. The left and right insets show the thickness profiles $h(x)$ corresponding to the symmetric and asymmetric states at $\Gamma=10^{-3}$.}
\label{fig:biffull}
\end{center}
\end{figure}

\section{Results and discussion}\label{sec:results}

\subsection{Free drops}

Fixed points of the dynamical system \eqref{EOM} clearly satisfy $\mathcal{P}=\mathcal{T}=\mathcal{F}=0$; equivalently, $\alpha=0$ and $\mathcal{U}=\Omega R$ (``pure-rolling''). In particular, there is always the trivial fixed point $(\mathcal{U},\Omega)=(0,0)$ which corresponds to the usual stationary-symmetric state of a Leidenfrost drop. We find that this symmetric state, however, is stable only for $\Gamma>\Gamma_F\approx0.0235$.
For $\Gamma<\Gamma_F$, there are left- and right-going pure-rolling steady states (corresponding to non-trivial zeros of $\alpha$ as an odd function of $\mathcal{U}+\Omega R = 2\mathcal{U}$, which bifurcate at $\Gamma_F$ from the trivial zero). This supercritical pitchfork bifurcation is shown in Fig.~\ref{fig:biffull}. The insets show the thickness profiles ${h}(x)$ for the symmetric and asymmetric steady states at $\Gamma=10^{-3}$.

\subsection{Immobilized drops}
  
It is clear that steady pure-rolling states do not conform to the constant-acceleration motion observed in the experiments \cite{Bouillant:18}. It appears important that in the experiments the drop is initially immobilized, by a needle, and then released; while the drop is immobilized, a rotational flow builds up within the drop.

\begin{figure}[t!]
\begin{center}
\includegraphics[scale=0.45]{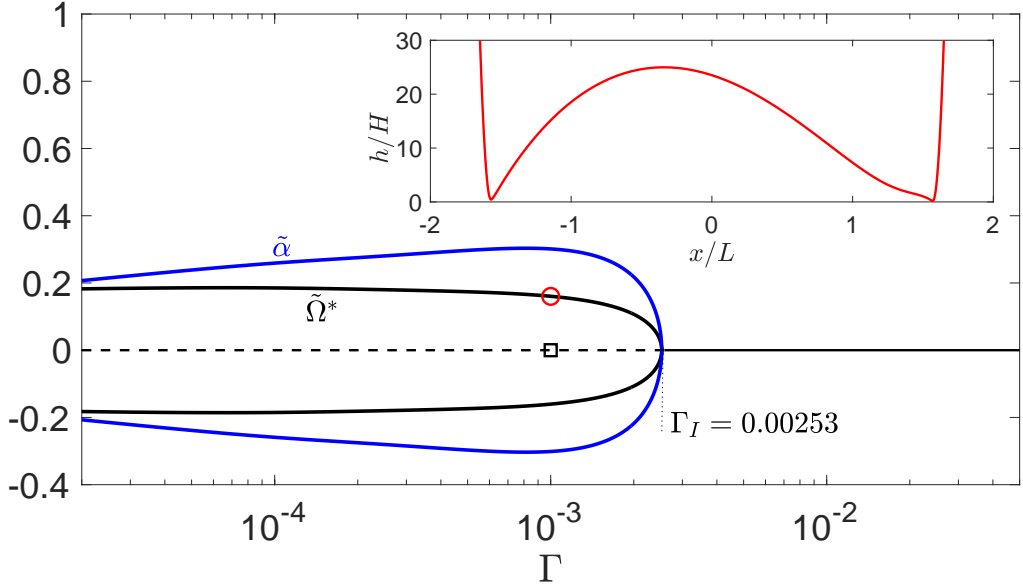}
\caption{Bifurcation diagram for an immobilized drop, where we define the normalized angular speed $\tilde{\Omega}^*(\Gamma)= \Omega^* R/U$ and rescaled slope $\tilde{\alpha} = \alpha g T/U$, which is also the initial acceleration of an immobilized drop released from its steady rotational state normalized by $U/T$. The inset shows the thickness profile $h(x)$ corresponding to the rotational state at $\Gamma=10^{-3}$.}
\label{fig:bifreduced}
\end{center}
\end{figure}

To model this first stage in the experiment, where the drop is immobilized, we add to (\ref{EOM}a) a horizontal constraint force. If we assume that this force passes through the drop center of mass, then the angular-momentum balance (\ref{EOM}b) remains unchanged, only that the forces and torques are calculated from the lubrication problem with $\mathcal{U}=0$; we thereby find that an immobilized drop is governed by a reduced one-dimensional dynamical system for its angular velocity $\Omega$.
Analogously to the free-drop case, we find that the symmetric steady state $\Omega=0$ of this reduced system is stable for $\Gamma>\Gamma_I\approx 0.00253$. For $\Gamma<\Gamma_I$, there are stable steady states of constant angular velocity ${\Omega}={\Omega}^*$ satisfying 
\begin{equation}\label{reduced bif}
\mathcal{P}-\mathcal{F}=\mathcal{T}/R.
\end{equation} 
This supercritical pitchfork bifurcation is shown in Fig.~\ref{fig:bifreduced}, the inset depicting the thickness profile ${h}(x)$ for the rotational steady state at $\Gamma=10^{-3}$. 

\subsection{Released drops}

\begin{figure}[t!]
\begin{center}
\includegraphics[scale=0.4]{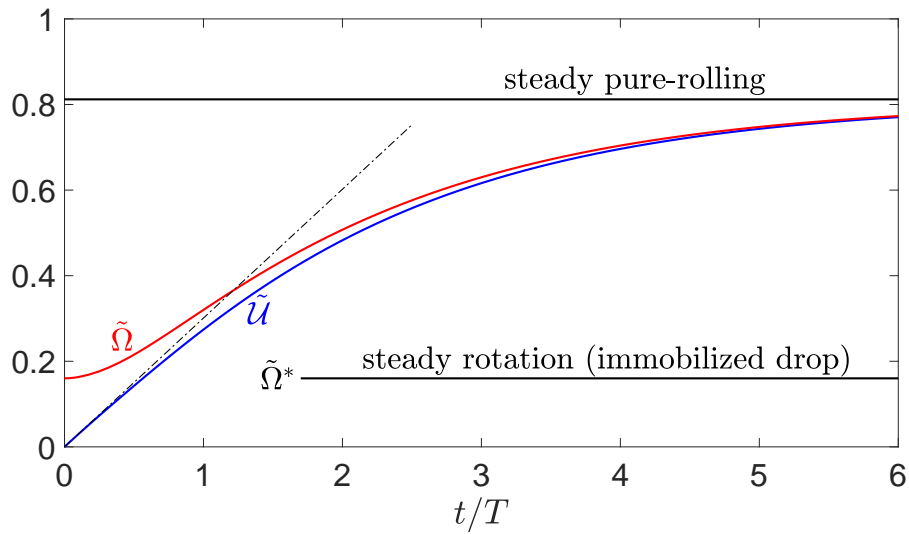}
\caption{
Time evolution of $\tilde{\mathcal{U}}=\mathcal{U}/U$ and $\tilde{\Omega} = \Omega R/U$ for a drop released at time $t=0$ from an immobilized steady state of constant angular velocity $\Omega^*$, for $\Gamma=10^{-3}$. The dashed line depicts the initial acceleration of the drop \eqref{acc} normalized by $U/T$.}
\label{fig:time}
\end{center}
\end{figure}

Now assume that an immobilized drop rotating at the stationary angular velocity $\Omega={\Omega}^*$ (or $-\Omega^*$) is suddenly set free at $t=0$. From (\ref{EOM}a) and \eqref{reduced bif}, the drop starts to move with acceleration $d\mathcal{U}/dt=\mathcal{T}/(mR)$. Using (\ref{PT in h}b), this initial acceleration is written 
\begin{equation}\label{acc}
\frac{d\mathcal{U}}{dt}= \alpha g.
\end{equation}
A simulation of the free-drop equations \eqref{EOM}, starting from this initial state, is shown in Fig.~\ref{fig:time}. For $t\ll T$, the acceleration and angular velocity remain approximately constant. We argue that this initial period corresponds to the observed regime  \footnote{From private communications with the authors of \cite{Bouillant:18}, $\mathcal{U}<\Omega R$ throughout the measurements; moreover, the experimental data suggests an inertial time scale $\approx 10$ times larger than the observation time.}. Theoretically, at large times the drop approaches the steady pure-rolling state.

The correlation \eqref{acc} is precisely the one found in the experiments \cite{Bouillant:18} if we identify the characteristic slope $\alpha$ with the measured ``inclination angle'', namely the geometric angle between the surface and a line passing through the local neck minima (numerically, $\alpha$ is within  
$14\%$ of that geometric angle for $\Gamma<\Gamma_I$). 
The acceleration $\alpha g$, rescaled by $U/T$, is plotted in Fig.~\ref{fig:bifreduced} as a function of $\Gamma$. This function exhibits a bifurcation from zero at $\Gamma=\Gamma_I$. The maximum of this function implies a maximum of $\alpha$ (and the acceleration) as a function of $R$, with all other parameters fixed; this maximum occurs at a value of $R$ increasing with the temperature difference $\Delta T$. 

On the one hand, all of the above trends are in qualitative agreement with the experiments.
On the other hand, any quantitative agreement would be fortuitous, given that our model is two-dimensional. Thus, for the physical parameters in the experiments, the value $\Gamma_I$ suggests a bifurcation radius $\approx 4$mm which overestimates the observed value $\approx 0.6$mm and appears to contradict our assumption $B\ll1$ (the capillary length is $\approx 2.5$mm for the experimental parameters); that the experimental  bifurcation radius satisfies the small-$B$ condition, however, suggests that that assumption is in fact physically representative.

We note that in \cite{Bouillant:18} the force $\alpha mg$ is interpreted as a propulsion force given by the horizontal projection of the lift force \eqref{vforce}. This heuristic argument, however, ignores shear stresses, which halve the propulsion, i.e., $\mathcal{P}=\alpha mg/2$, and generate friction $\mathcal{F}$. Only in the steady rotational state \eqref{reduced bif} of an immobilized drop it happens to be that $\mathcal{P}-\mathcal{F}=\alpha mg$.

\section{Concluding remarks}\label{sec:conc}
We have presented a 
two-dimensional model of small Leidenfrost drops which predicts symmetry-breaking spontaneous dynamics. The model couples the rigid-body equations of motion of the drop and an instantaneous lubrication model of the vapor thin film; the equations of motion output the drop's linear and angular velocities, which are continuously translated by the lubrication model, wherein evaporation acts like nonlinear gain, into forces and torques on the drop that are fed back into the equations of motion as inputs. The stability of this feedback loop is determined by the single  dimensionless parameter \eqref{Gamma def}. 

Although our model is two-dimensional, it qualitatively agrees with the three-dimensional experiments in \cite{Bouillant:18}. Our model, in turn, illuminates the mechanics underlying the symmetry-breaking instability and clarifies the propulsion mechanism in the symmetry-broken regime. Furthermore, our model links the constant-acceleration motion observed in \cite{Bouillant:18} to the experimental ``initial conditions,''
while unraveling alternative dynamical regimes, including ``pure-rolling'' steady states, not yet observed experimentally.

Generalizing our model to three dimensions would hopefully allow  \emph{quantitative} agreement with experiments. It would also be desirable to generalize the theory to moderate $B$, where the experiments show a second bifurcation from the rotational state to a symmetric state characterized by a two-cell drop-scale flow. We suspect this latter transition occurs as the liquid flow shear-driven by the vapor film, which for $B\ll1$ is localized and negligible in magnitude relative to the vapor and inertial drop-scale flows, extends to the whole drop and grows comparable in magnitude to the vapor flow.

Lastly, we note that symmetry-breaking spontaneous motion was recently observed also for so-called inverse-Leidenfrost drops, namely drops levitating above an evaporating liquid bath \cite{Gauthier:19}. In those experiments, however, the drop does not roll and the bath-vapor interface, rather than the drop-vapor interface, significantly deforms. Owing to these differences, our model cannot be directly applied to that problem. Furthermore, while the two-dimensional numerical simulations in \cite{Gauthier:19} exhibit symmetry breaking, the analytical modeling therein assumes, rather than describes, the observed asymmetry. Thus, a theory of symmetry breaking in the inverse-Leidenfrost scenario remains an open problem.

\begin{acknowledgments}
We are grateful to Ambre Bouillant and David Qu\'er\'e for clarifications on \cite{Bouillant:18}.
\end{acknowledgments}

\bibliography{refs}

\begin{thebibliography}{21}%
\makeatletter
\providecommand \@ifxundefined [1]{%
 \@ifx{#1\undefined}
}%
\providecommand \@ifnum [1]{%
 \ifnum #1\expandafter \@firstoftwo
 \else \expandafter \@secondoftwo
 \fi
}%
\providecommand \@ifx [1]{%
 \ifx #1\expandafter \@firstoftwo
 \else \expandafter \@secondoftwo
 \fi
}%
\providecommand \natexlab [1]{#1}%
\providecommand \enquote  [1]{``#1''}%
\providecommand \bibnamefont  [1]{#1}%
\providecommand \bibfnamefont [1]{#1}%
\providecommand \citenamefont [1]{#1}%
\providecommand \href@noop [0]{\@secondoftwo}%
\providecommand \href [0]{\begingroup \@sanitize@url \@href}%
\providecommand \@href[1]{\@@startlink{#1}\@@href}%
\providecommand \@@href[1]{\endgroup#1\@@endlink}%
\providecommand \@sanitize@url [0]{\catcode `\\12\catcode `\$12\catcode
  `\&12\catcode `\#12\catcode `\^12\catcode `\_12\catcode `\%12\relax}%
\providecommand \@@startlink[1]{}%
\providecommand \@@endlink[0]{}%
\providecommand \url  [0]{\begingroup\@sanitize@url \@url }%
\providecommand \@url [1]{\endgroup\@href {#1}{\urlprefix }}%
\providecommand \urlprefix  [0]{URL }%
\providecommand \Eprint [0]{\href }%
\providecommand \doibase [0]{http://dx.doi.org/}%
\providecommand \selectlanguage [0]{\@gobble}%
\providecommand \bibinfo  [0]{\@secondoftwo}%
\providecommand \bibfield  [0]{\@secondoftwo}%
\providecommand \translation [1]{[#1]}%
\providecommand \BibitemOpen [0]{}%
\providecommand \bibitemStop [0]{}%
\providecommand \bibitemNoStop [0]{.\EOS\space}%
\providecommand \EOS [0]{\spacefactor3000\relax}%
\providecommand \BibitemShut  [1]{\csname bibitem#1\endcsname}%
\let\auto@bib@innerbib\@empty
%</preamble>
\bibitem [{\citenamefont {Biance}\ \emph {et~al.}(2003)\citenamefont {Biance},
  \citenamefont {Clanet},\ and\ \citenamefont {Qu{\'e}r{\'e}}}]{Biance:03}%
  \BibitemOpen
  \bibfield  {author} {\bibinfo {author} {\bibfnamefont {A.-L.}\ \bibnamefont
  {Biance}}, \bibinfo {author} {\bibfnamefont {C.}~\bibnamefont {Clanet}}, \
  and\ \bibinfo {author} {\bibfnamefont {D.}~\bibnamefont {Qu{\'e}r{\'e}}},\
  }\bibfield  {title} {\enquote {\bibinfo {title} {{Leidenfrost} drops},}\
  }\href@noop {} {\bibfield  {journal} {\bibinfo  {journal} {Phys. Fluids}\
  }\textbf {\bibinfo {volume} {15}},\ \bibinfo {pages} {1632--1637} (\bibinfo
  {year} {2003})}\BibitemShut {NoStop}%
\bibitem [{\citenamefont {Qu{\'e}r{\'e}}(2013)}]{Quere:13}%
  \BibitemOpen
  \bibfield  {author} {\bibinfo {author} {\bibfnamefont {D.}~\bibnamefont
  {Qu{\'e}r{\'e}}},\ }\bibfield  {title} {\enquote {\bibinfo {title}
  {{Leidenfrost} dynamics},}\ }\href@noop {} {\bibfield  {journal} {\bibinfo
  {journal} {Annu. Rev. Fluid Mech.}\ }\textbf {\bibinfo {volume} {45}},\
  \bibinfo {pages} {197--215} (\bibinfo {year} {2013})}\BibitemShut {NoStop}%
\bibitem [{\citenamefont {Bouillant}\ \emph
  {et~al.}(2018{\natexlab{a}})\citenamefont {Bouillant}, \citenamefont
  {Mouterde}, \citenamefont {Bourrianne}, \citenamefont {Lagarde},
  \citenamefont {Clanet},\ and\ \citenamefont {Qu{\'e}r{\'e}}}]{Bouillant:18}%
  \BibitemOpen
  \bibfield  {author} {\bibinfo {author} {\bibfnamefont {A.}~\bibnamefont
  {Bouillant}}, \bibinfo {author} {\bibfnamefont {T.}~\bibnamefont {Mouterde}},
  \bibinfo {author} {\bibfnamefont {P.}~\bibnamefont {Bourrianne}}, \bibinfo
  {author} {\bibfnamefont {A.}~\bibnamefont {Lagarde}}, \bibinfo {author}
  {\bibfnamefont {C.}~\bibnamefont {Clanet}}, \ and\ \bibinfo {author}
  {\bibfnamefont {D.}~\bibnamefont {Qu{\'e}r{\'e}}},\ }\bibfield  {title}
  {\enquote {\bibinfo {title} {Leidenfrost wheels},}\ }\href@noop {} {\bibfield
   {journal} {\bibinfo  {journal} {Nat. Phys}\ }\textbf {\bibinfo {volume}
  {14}},\ \bibinfo {pages} {1188--1192} (\bibinfo {year}
  {2018}{\natexlab{a}})}\BibitemShut {NoStop}%
\bibitem [{\citenamefont {Bouillant}\ \emph
  {et~al.}(2018{\natexlab{b}})\citenamefont {Bouillant}, \citenamefont
  {Mouterde}, \citenamefont {Bourrianne}, \citenamefont {Clanet},\ and\
  \citenamefont {Qu{\'e}r{\'e}}}]{Bouillant:18:gallery}%
  \BibitemOpen
  \bibfield  {author} {\bibinfo {author} {\bibfnamefont {A.}~\bibnamefont
  {Bouillant}}, \bibinfo {author} {\bibfnamefont {T.}~\bibnamefont {Mouterde}},
  \bibinfo {author} {\bibfnamefont {P.}~\bibnamefont {Bourrianne}}, \bibinfo
  {author} {\bibfnamefont {C.}~\bibnamefont {Clanet}}, \ and\ \bibinfo {author}
  {\bibfnamefont {D.}~\bibnamefont {Qu{\'e}r{\'e}}},\ }\bibfield  {title}
  {\enquote {\bibinfo {title} {Symmetry breaking in {Leidenfrost} flows},}\
  }\href@noop {} {\bibfield  {journal} {\bibinfo  {journal} {Phys. Rev.
  Fluids}\ }\textbf {\bibinfo {volume} {3}},\ \bibinfo {pages} {100502}
  (\bibinfo {year} {2018}{\natexlab{b}})}\BibitemShut {NoStop}%
\bibitem [{\citenamefont {Miller}(2018)}]{Miller:18}%
  \BibitemOpen
  \bibfield  {author} {\bibinfo {author} {\bibfnamefont {J.~L.}\ \bibnamefont
  {Miller}},\ }\bibfield  {title} {\enquote {\bibinfo {title} {{Leidenfrost}
  drops are on a roll},}\ }\href@noop {} {\bibfield  {journal} {\bibinfo
  {journal} {Phys. Today}\ }\textbf {\bibinfo {volume} {71}},\ \bibinfo {pages}
  {14--15} (\bibinfo {year} {2018})}\BibitemShut {NoStop}%
\bibitem [{\citenamefont {Celestini}\ and\ \citenamefont
  {Kirstetter}(2012)}]{Celestini2:12}%
  \BibitemOpen
  \bibfield  {author} {\bibinfo {author} {\bibfnamefont {F.}~\bibnamefont
  {Celestini}}\ and\ \bibinfo {author} {\bibfnamefont {G.}~\bibnamefont
  {Kirstetter}},\ }\bibfield  {title} {\enquote {\bibinfo {title} {Effect of an
  electric field on a {Leidenfrost} droplet},}\ }\href@noop {} {\bibfield
  {journal} {\bibinfo  {journal} {Soft Matter}\ }\textbf {\bibinfo {volume}
  {8}},\ \bibinfo {pages} {5992--5995} (\bibinfo {year} {2012})}\BibitemShut
  {NoStop}%
\bibitem [{\citenamefont {Burton}\ \emph {et~al.}(2012)\citenamefont {Burton},
  \citenamefont {Sharpe}, \citenamefont {Van Der~Veen}, \citenamefont
  {Franco},\ and\ \citenamefont {Nagel}}]{Burton:12}%
  \BibitemOpen
  \bibfield  {author} {\bibinfo {author} {\bibfnamefont {J.~C.}\ \bibnamefont
  {Burton}}, \bibinfo {author} {\bibfnamefont {A.~L.}\ \bibnamefont {Sharpe}},
  \bibinfo {author} {\bibfnamefont {R.~C.~A.}\ \bibnamefont {Van Der~Veen}},
  \bibinfo {author} {\bibfnamefont {A.}~\bibnamefont {Franco}}, \ and\ \bibinfo
  {author} {\bibfnamefont {S.~R.}\ \bibnamefont {Nagel}},\ }\bibfield  {title}
  {\enquote {\bibinfo {title} {Geometry of the vapor layer under a
  {Leidenfrost} drop},}\ }\href@noop {} {\bibfield  {journal} {\bibinfo
  {journal} {Phys. Rev. Lett.}\ }\textbf {\bibinfo {volume} {109}},\ \bibinfo
  {pages} {074301} (\bibinfo {year} {2012})}\BibitemShut {NoStop}%
\bibitem [{\citenamefont {Pomeau}\ \emph {et~al.}(2012)\citenamefont {Pomeau},
  \citenamefont {Le~Berre}, \citenamefont {Celestini},\ and\ \citenamefont
  {Frisch}}]{Pomeau:12}%
  \BibitemOpen
  \bibfield  {author} {\bibinfo {author} {\bibfnamefont {Y.}~\bibnamefont
  {Pomeau}}, \bibinfo {author} {\bibfnamefont {M.}~\bibnamefont {Le~Berre}},
  \bibinfo {author} {\bibfnamefont {F.}~\bibnamefont {Celestini}}, \ and\
  \bibinfo {author} {\bibfnamefont {T.}~\bibnamefont {Frisch}},\ }\bibfield
  {title} {\enquote {\bibinfo {title} {The {Leidenfrost} effect: From
  quasi-spherical droplets to puddles},}\ }\href@noop {} {\bibfield  {journal}
  {\bibinfo  {journal} {C. R. Mec.}\ }\textbf {\bibinfo {volume} {340}},\
  \bibinfo {pages} {867--881} (\bibinfo {year} {2012})}\BibitemShut {NoStop}%
\bibitem [{\citenamefont {Sobac}\ \emph {et~al.}(2014)\citenamefont {Sobac},
  \citenamefont {Rednikov}, \citenamefont {Dorbolo},\ and\ \citenamefont
  {Colinet}}]{Sobac:14}%
  \BibitemOpen
  \bibfield  {author} {\bibinfo {author} {\bibfnamefont {B.}~\bibnamefont
  {Sobac}}, \bibinfo {author} {\bibfnamefont {A.}~\bibnamefont {Rednikov}},
  \bibinfo {author} {\bibfnamefont {S.}~\bibnamefont {Dorbolo}}, \ and\
  \bibinfo {author} {\bibfnamefont {P.}~\bibnamefont {Colinet}},\ }\bibfield
  {title} {\enquote {\bibinfo {title} {{Leidenfrost} effect: Accurate drop
  shape modeling and refined scaling laws},}\ }\href@noop {} {\bibfield
  {journal} {\bibinfo  {journal} {Phys. Rev. E}\ }\textbf {\bibinfo {volume}
  {90}},\ \bibinfo {pages} {053011} (\bibinfo {year} {2014})}\BibitemShut
  {NoStop}%
\bibitem [{\citenamefont {Sobac}\ \emph {et~al.}(2017)\citenamefont {Sobac},
  \citenamefont {Rednikov}, \citenamefont {Dorbolo},\ and\ \citenamefont
  {Colinet}}]{Sobac:17}%
  \BibitemOpen
  \bibfield  {author} {\bibinfo {author} {\bibfnamefont {B.}~\bibnamefont
  {Sobac}}, \bibinfo {author} {\bibfnamefont {A.}~\bibnamefont {Rednikov}},
  \bibinfo {author} {\bibfnamefont {S.}~\bibnamefont {Dorbolo}}, \ and\
  \bibinfo {author} {\bibfnamefont {P.}~\bibnamefont {Colinet}},\ }\bibfield
  {title} {\enquote {\bibinfo {title} {Self-propelled {Leidenfrost} drops on a
  thermal gradient: A theoretical study},}\ }\href@noop {} {\bibfield
  {journal} {\bibinfo  {journal} {Phys. Fluids}\ }\textbf {\bibinfo {volume}
  {29}},\ \bibinfo {pages} {082101} (\bibinfo {year} {2017})}\BibitemShut
  {NoStop}%
\bibitem [{\citenamefont {Duchemin}\ \emph {et~al.}(2005)\citenamefont
  {Duchemin}, \citenamefont {Lister},\ and\ \citenamefont
  {Lange}}]{Duchemin:05}%
  \BibitemOpen
  \bibfield  {author} {\bibinfo {author} {\bibfnamefont {L.}~\bibnamefont
  {Duchemin}}, \bibinfo {author} {\bibfnamefont {J.~R.}\ \bibnamefont
  {Lister}}, \ and\ \bibinfo {author} {\bibfnamefont {U.}~\bibnamefont
  {Lange}},\ }\bibfield  {title} {\enquote {\bibinfo {title} {Static shapes of
  levitated viscous drops},}\ }\href@noop {} {\bibfield  {journal} {\bibinfo
  {journal} {J. Fluid Mech.}\ }\textbf {\bibinfo {volume} {533}},\ \bibinfo
  {pages} {161--170} (\bibinfo {year} {2005})}\BibitemShut {NoStop}%
\bibitem [{\citenamefont {Snoeijer}\ \emph {et~al.}(2009)\citenamefont
  {Snoeijer}, \citenamefont {Brunet},\ and\ \citenamefont
  {Eggers}}]{Snoeijer:09}%
  \BibitemOpen
  \bibfield  {author} {\bibinfo {author} {\bibfnamefont {J.~H.}\ \bibnamefont
  {Snoeijer}}, \bibinfo {author} {\bibfnamefont {P.}~\bibnamefont {Brunet}}, \
  and\ \bibinfo {author} {\bibfnamefont {J.}~\bibnamefont {Eggers}},\
  }\bibfield  {title} {\enquote {\bibinfo {title} {Maximum size of drops
  levitated by an air cushion},}\ }\href@noop {} {\bibfield  {journal}
  {\bibinfo  {journal} {Phys. Rev. E}\ }\textbf {\bibinfo {volume} {79}},\
  \bibinfo {pages} {036307} (\bibinfo {year} {2009})}\BibitemShut {NoStop}%
\bibitem [{\citenamefont {Mahadevan}\ and\ \citenamefont
  {Pomeau}(1999)}]{Mahadevan:99}%
  \BibitemOpen
  \bibfield  {author} {\bibinfo {author} {\bibfnamefont {L.}~\bibnamefont
  {Mahadevan}}\ and\ \bibinfo {author} {\bibfnamefont {Y.}~\bibnamefont
  {Pomeau}},\ }\bibfield  {title} {\enquote {\bibinfo {title} {Rolling
  droplets},}\ }\href@noop {} {\bibfield  {journal} {\bibinfo  {journal} {Phys.
  Fluids}\ }\textbf {\bibinfo {volume} {11}},\ \bibinfo {pages} {2449--2453}
  (\bibinfo {year} {1999})}\BibitemShut {NoStop}%
\bibitem [{\citenamefont {Yariv}\ and\ \citenamefont
  {Schnitzer}(2019)}]{Yariv:19:2Drolling}%
  \BibitemOpen
  \bibfield  {author} {\bibinfo {author} {\bibfnamefont {E.}~\bibnamefont
  {Yariv}}\ and\ \bibinfo {author} {\bibfnamefont {O.}~\bibnamefont
  {Schnitzer}},\ }\bibfield  {title} {\enquote {\bibinfo {title} {Speed of
  rolling droplets},}\ }\href@noop {} {\bibfield  {journal} {\bibinfo
  {journal} {Phys. Rev. Fluids}\ }\textbf {\bibinfo {volume} {4}},\ \bibinfo
  {pages} {093602} (\bibinfo {year} {2019})}\BibitemShut {NoStop}%
\bibitem [{\citenamefont {Schnitzer}\ \emph {et~al.}(in press)\citenamefont
  {Schnitzer}, \citenamefont {Davis},\ and\ \citenamefont
  {Yariv}}]{Schnitzer:20}%
  \BibitemOpen
  \bibfield  {author} {\bibinfo {author} {\bibfnamefont {O.}~\bibnamefont
  {Schnitzer}}, \bibinfo {author} {\bibfnamefont {A.~M.~J.}\ \bibnamefont
  {Davis}}, \ and\ \bibinfo {author} {\bibfnamefont {E.}~\bibnamefont
  {Yariv}},\ }\bibfield  {title} {\enquote {\bibinfo {title} {Rolling of
  nonwetting droplets down a gently inclined plane},}\ }\href@noop {}
  {\bibfield  {journal} {\bibinfo  {journal} {J. Fluid Mech.}\ } (\bibinfo
  {year} {in press})}\BibitemShut {NoStop}%
\bibitem [{Note1()}]{Note1}%
  \BibitemOpen
  \bibinfo {note} {Relative to the capillary stress $\gamma /R$, the inertial
  and viscous dynamic stresses in the liquid respectively scale like the Weber
  number $\protect \mathrm {We}=\bar {\rho } U^2 R/\gamma $ and $B\protect
  \mathrm {Ca}$, wherein $\protect \mathrm {Ca}=\bar {\mu } U/\gamma $ is the
  capillary number and $U$ the characteristic velocity defined in \protect \S
  \protect \S \ref {ssec:scalings}. (The $B$ factor is because the viscous
  stress vanishes for a rigid-body motion.) We require these dynamic stresses
  to be $o(B\gamma /R)$, rather than $o(\gamma /R)$, for the reason mentioned
  below \protect \textup {\hbox {\mathsurround \z@ \protect \normalfont
  (\ignorespaces \ref {vforce}\unskip \@@italiccorr )}}. Furthermore, the
  condition $1/\Omega \ll T$, or $R/U \ll T$, is equivalent to $\Gamma B^2\ll
  \protect \mathrm {We}$, wherein $\Gamma $ is defined in \protect \textup
  {\hbox {\mathsurround \z@ \protect \normalfont (\ignorespaces \ref {Gamma
  def}\unskip \@@italiccorr )}} and we use the estimates for $U$ and $T$
  derived in \protect \S \protect \S \ref {ssec:scalings}. We accordingly
  assume that $\Gamma B^2\ll \protect \mathrm {We}\ll B$ and $\protect \mathrm
  {Ca}\ll 1$.}\BibitemShut {Stop}%
\bibitem [{Note2()}]{Note2}%
  \BibitemOpen
  \bibinfo {note} {The deviation of the drop shape from a circular shape, due
  to gravity, implies a negligible $O(B\Omega R)$ velocity perturbation \cite
  {Yariv:19:2Drolling}. Meanwhile, a tangential-stress balance at the
  liquid-vapor interface implies a velocity perturbation $O(\mu L/\bar {\mu
  }H)$ relative to the characteristic vapor velocity $U$, where $H=\Gamma B^2R$
  and $\Gamma $ is defined in \protect \textup {\hbox {\mathsurround \z@
  \protect \normalfont (\ignorespaces \ref {Gamma def}\unskip \@@italiccorr
  )}}. For $U=O(\Omega R)$ this implies the condition $\mu /\bar {\mu }\ll
  B\Gamma $. Flow measurements at the drop base confirm an approximately
  uniform flow \cite {Bouillant:18}.}\BibitemShut {Stop}%
\bibitem [{Note3()}]{Note3}%
  \BibitemOpen
  \bibinfo {note} {In the co-moving frame, $h$ varies on the inertial time
  scale $T$. Using the estimates derived in \protect \S \protect \S \ref
  {ssec:scalings}, the ratio between $d{h}/dt$ and the normal vapor speed is $B
  R/U T$, which is small since $B\ll 1$ and $T \gg R/U$.}\BibitemShut {Stop}%
\bibitem [{Note4()}]{Note4}%
  \BibitemOpen
  \bibinfo {note} {See Supplemental Material for a dimensionless formulation of
  the model.}\BibitemShut {Stop}%
\bibitem [{Note5()}]{Note5}%
  \BibitemOpen
  \bibinfo {note} {From private communications with the authors of \cite
  {Bouillant:18}, $\protect \mathcal {U}<\Omega R$ throughout the measurements;
  moreover, the experimental data suggests an inertial time scale $\approx 10$
  times larger than the observation time.}\BibitemShut {Stop}%
\bibitem [{\citenamefont {Gauthier}\ \emph {et~al.}(2019)\citenamefont
  {Gauthier}, \citenamefont {Diddens}, \citenamefont {Proville}, \citenamefont
  {Lohse},\ and\ \citenamefont {van~der Meer}}]{Gauthier:19}%
  \BibitemOpen
  \bibfield  {author} {\bibinfo {author} {\bibfnamefont {A.}~\bibnamefont
  {Gauthier}}, \bibinfo {author} {\bibfnamefont {C.}~\bibnamefont {Diddens}},
  \bibinfo {author} {\bibfnamefont {R.}~\bibnamefont {Proville}}, \bibinfo
  {author} {\bibfnamefont {D.}~\bibnamefont {Lohse}}, \ and\ \bibinfo {author}
  {\bibfnamefont {D.}~\bibnamefont {van~der Meer}},\ }\bibfield  {title}
  {\enquote {\bibinfo {title} {Self-propulsion of inverse {Leidenfrost} drops
  on a cryogenic bath},}\ }\href@noop {} {\bibfield  {journal} {\bibinfo
  {journal} {Proc. Natl. Acad.}\ }\textbf {\bibinfo {volume} {116}},\ \bibinfo
  {pages} {1174--1179} (\bibinfo {year} {2019})}\BibitemShut {NoStop}%
\end{thebibliography}%
\end{document}